\documentclass{article}

\usepackage{arxiv}

\usepackage[utf8]{inputenc} % allow utf-8 input
\usepackage[T1]{fontenc}    % use 8-bit T1 fonts
\usepackage{hyperref}       % hyperlinks
\usepackage{url}            % simple URL typesetting
\usepackage{booktabs}       % professional-quality tables
\usepackage{amsfonts}       % blackboard math symbols
\usepackage{nicefrac}       % compact symbols for 1/2, etc.
\usepackage{microtype}      % microtypography
\usepackage{lipsum}
\usepackage{graphicx}
\usepackage{multirow}
\usepackage{xtab}
\usepackage{longtable}
 \usepackage{xcolor}

\definecolor{adapt}{HTML}{4e79a7}
\definecolor{cohesion}{HTML}{f28e2b}
\definecolor{coregulate}{HTML}{59a14f}
\definecolor{emotion}{HTML}{e15759}
\definecolor{explore}{HTML}{af7aa1}
\definecolor{monitor}{HTML}{edc948}
\definecolor{plan}{HTML}{76b7b2}
\definecolor{synthesis}{HTML}{ff9da7}
\definecolor{understand}{HTML}{9c755f}

\definecolor{yellowcom}{HTML}{9eb022}
\definecolor{orangecom}{HTML}{fd5308}
\definecolor{greencom}{HTML}{68b034}

\definecolor{planningcluster}{HTML}{e1575a}
\definecolor{consolidationcluster}{HTML}{58a14e}
\definecolor{explorationcluster}{HTML}{6691c0}
\definecolor{socialcluster}{HTML}{f28e2a}

\title {Transition Network Analysis: A Novel Framework for Modeling, Visualizing, and Identifying the Temporal Patterns of Learners and Learning Processes}

\begin{document}
\author{
Mohammed Saqr \\
 University of Eastern Finland \\
 Joensuu, Finland \\
\texttt{mohammed.saqr@uef.fi} \\
 \And
Sonsoles López-Pernas \\
 University of Eastern Finland \\
 Joensuu, Finland \\
\texttt{sonsoles.lopez@uef.fi} \\
 \And
Tiina Törmänen \\
 University of Oulu \\
 Oulu, Finland \\
\texttt{Tiina.Tormanen@oulu.fi} \\
 \And
Rogers Kaliisa \\
 University of Oslo \\
 Oslo, Norway \\
\texttt{rogers.kaliisa@iped.uio.no} \\
 \And
Kamila Misiejuk \\
 FernUniversität in Hagen \\
 Hagen, Germany \\
 University of Bergen \\
 Bergen, Norway \\
\texttt{kamila.misiejuk@fernuni-hagen.de} \\
 \And
Santtu Tikka \\
 University of Jyväskylä \\
 Jyväskylä, Finland \\
\texttt{santtu.tikka@jyu.fi} \\
}

\rhead{\textit{Transition Network Analysis} - Saqr et al. 2025}

\maketitle


\begin{abstract}
This paper presents a novel learning analytics method: Transition Network Analysis (TNA), a method that integrates Stochastic Process Mining and probabilistic graph representation to model, visualize, and identify transition patterns in the learning process data. Combining the relational and temporal aspects into a single lens offers capabilities beyond either framework, including centralities to capture important learning events, community detection to identify behavior patterns, and clustering to reveal temporal patterns. Furthermore, TNA introduces several significance tests that go beyond either method and add rigor to the analysis. Here, we introduce the theoretical and mathematical foundations of TNA and we demonstrate the functionalities of TNA with a case study where students (n=191) engaged in small-group collaboration to map patterns of group dynamics using the theories of co-regulation and socially-shared regulated learning. The analysis revealed that TNA can map the regulatory processes as well as identify important events, patterns, and clusters. Bootstrap validation established the significant transitions and eliminated spurious transitions. As such, TNA can capture learning dynamics and provide a robust framework for investigating the temporal evolution of learning processes. Future directions include ---inter alia--- expanding estimation methods, reliability assessment, and building longitudinal TNA.

\end{abstract}



%\received{2024}
%received[revised]{2024}
%\received[accepted]{2024}

\maketitle

\section{Introduction}

The significance of studying the learning process lies in the fact that
learning cannot be directly observed but rather manifests itself in ``the very actions that learners perform'' and how these actions unfold over time
\cite[p.~269]{Winne_2010}. These events
may be more indicative of learning than their descriptions and
potentially more accurate than recollections or intentions of learning
\cite{Gašević_Jovanović_Pardo_Dawson_2017}.
A significant body of literature has embraced such a view and applied a
variety of lenses to chart the evolution of learning
processes
\cite{Malmberg_Järvelä_Järvenoja_2017,Matcha_Gašević_Ahmad_Uzir_Jovanović_Pardo_Maldonado-Mahauad_Pérez-Sanagustín_2019,Reimann_2009,Saint_Gaševic_Matcha_Uzir_Pardo_2020}.
However, capturing the full breadth of learning processes requires a
multifaceted approach due to the plurality of temporal levels,
relational aspects, and interdependencies between learning events
\cite{Chen_Poquet_2020,Poquet_Dawson_Dowell_2017,Saint_Gaševic_Matcha_Uzir_Pardo_2020}.
Whereas several methods exist, approaches that capture ---and take 
full advantage of--- the relational and temporal dimensions of learning
data are needed. To address this issue, we introduce a novel
theory-informed learning analytics method to address
questions of significance for learning analytics and the learning
sciences at large called \textbf{Transition Network Analysis (TNA)}, a new
framework ---and software--- for analyzing and modeling the learning
processes. TNA integrates ---and builds on--- Stochastic Process Mining
(SPM) and probabilistic network (PN) models for graph representation and
analysis. The result is a modeling approach that not only takes
advantage of the capabilities of both methods but also offers several
functions beyond either framework alone.

In a nutshell, TNA models the transitions of learning events
---through Markov models--- as a graph and uses the
capabilities and potentials of graph representations to take advantage
of network analysis. Graph visualization offers a bird's
eye view of how the whole learning process unfolded. Furthermore,
researchers can discern transition patterns, e.g., dyads or cliques of common transitions by querying the matrix representation. Also, through graph representation, TNA offers a rich platform for computing
centrality measures that are of value to understand the dynamic transition process, e.g., in-strength (total incoming transitions) and betweenness (nodes that bridge transitions). Furthermore, TNA
networks can be clustered into temporal patterns and can model the
covariates that explain the emergence of such patterns. Community
finding can identify linked behaviors, which enables 
understanding of the commonalities of learning processes. In doing so, TNA
enables researchers to model, visualize, and identify patterns and
relationships between learning events. Most importantly, TNA introduces an unprecedented tool set of statistical inference methods to verify each modeling result. 
In the following, we describe the theoretical and mathematical foundations of TNA. We also present a case
study where we use TNA to study group regulation to showcase the
method\textquotesingle s potentials. Finally, we discuss TNA in view of
the existing methods, its limitations, and future directions. Please note that for brevity, we use the term TNA to refer to the method or the functions that have been developed within the framework and enabled by the methods.
\section{Background}
\subsection{Transition Network Analysis: Concept, theory, and
rationale}

Transitions between events have been commonly expressed as graphs of directed relationships between the events. These graphs are probabilistic networks where the nodes are events and the relations are ``direct probabilistic interactions between them''
\cite[p.~3]{Koller_Friedman_2009}. Recently, probabilistic network models have gained increasing attention ``to explore potential dynamic relationships
between'' observable variables and events leading to notable progress in psychology and social sciences in general \cite[p.~453]{Epskamp_Waldorp_Mõttus_Borsboom_2018}. Similar implementations can also be seen where
 transition processes are represented \cite[e.g.,]{Gales_Young_2008,Guerreiro_Silva_Amancio_2021,Silver_Silva_2021} or visualized \cite[e.g.,]{Helske_Helske_2019,Spedicato_2017}. In fact, similar to TNA, several Markovian network models have been proposed and empirically validated in other fields
under several labels, such as dynamic network models or Markov networks \cite{Nicolis2005,Zou2019,Vecchio_Miraglia_Maria_Rossini_2017}. 
Yet, an integrated framework that takes full advantage of the transition data, network analysis, and the mathematical potential of matrix representation (or graph theory) ---as is the case of TNA--- has not been endorsed or implemented in educational research. As such, we can conclude that utilizing a graph-theoretical representation of the transition process to model the relational dynamics of learning processes is timely and relevant. 

\begin{figure*}[!ht]
 \centering
 \includegraphics[width=\linewidth]{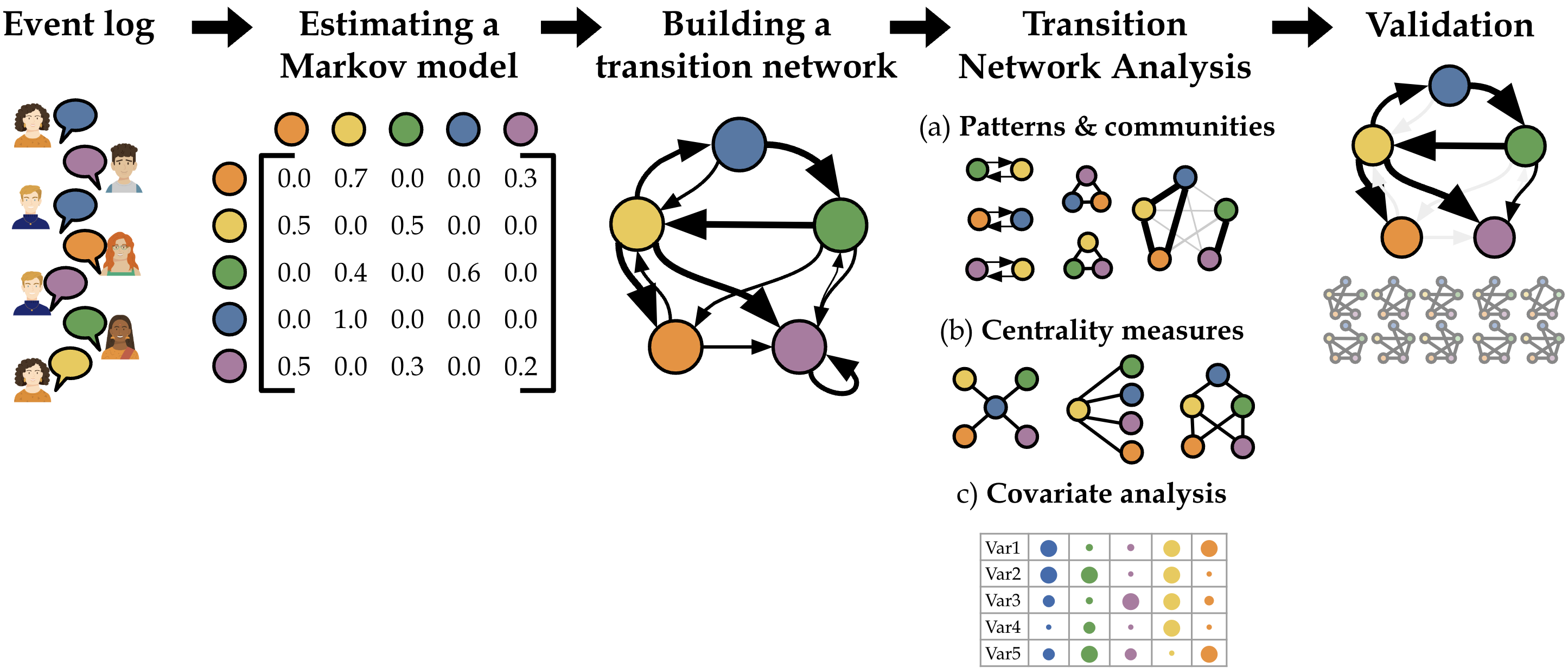}
 \caption{Outline of the TNA method: The transitions are captured from event logs with Markov models and represented as networks. TNA allows the discovery of different patterns, communities and clusters, as well as using covariates to explain these clusters. TNA models can also be tested for validity and stability using null models and bootstrapping.}
 \label{fig:tnamethod}
\end{figure*}

The building blocks of TNA are edges (or arcs) where the nodes are events, and the relationships are weighted directed transition probabilities from the events (see Figure \ref{fig:tnamethod}). A compilation of transition or a full representation of the process gives rise to an asymmetrical (directed) transition matrix which can be represented as a directed network \cite{Gales_Young_2008}. TNA utilizes statistical analysis and network methods, such as detecting patterns, community finding, calculating centrality measures, and conducting covariate analysis. Given that educational datasets may not be large enough, it is reasonable to assume that some edges may be spurious or ``noise''. To handle them, TNA introduces several techniques (e.g., bootstrapping and null models) to test for spurious transitions. In doing so, TNA modeling results in a robust and replicable model where each edge represents transition probability that differs significantly from random or from the null model \cite{Epskamp_Waldorp_Mõttus_Borsboom_2018, Mooney1993-az}. This
 stability and validity testing for every edge is a considerable advantage of TNA over existing methods (be it process mining or network analysis), where the validity of network or process models is rarely examined. More importantly, TNA robust process models offer a framework for hypothesis generation and testing.

\subsection{Theoretical underpinning}
The theoretical underpinning of TNA draws on the established view that
learning can be captured as an event-based process. Such a lens has been
applied ---and extended--- over the years through the work of several
researchers who applied a temporal lens to capture the sequences, the transitions, and the temporal unfolding of learning events and their typical patterns
\cite{López-Pernas_Misiejuk_Kaliisa_Conde-González_Saqr_2024,Malmberg_Järvelä_Järvenoja_2017,Matcha_Gašević_Ahmad_Uzir_Jovanović_Pardo_Maldonado-Mahauad_Pérez-Sanagustín_2019,Reimann_2009}.
In particular, TNA draws on Winne and Perry\textquotesingle s framework that the learning process unfolds as transitions (i.e., \emph{\textbf{contingencies)}} between \emph{\textbf{occurrences}} (the
actions learners take), such as from reading a passage to 
taking a note \cite{Winne_2010,Winne_Perry_2000}.
Repeated transitions form ---\emph{\textbf{patterns}}--- that can be inferred and ``computed using matrices that tally transitions from one SRL event to another'' \cite[p.~275]{Winne_2010}. TNA builds on ---and extends--- Winne's framework, given
that it encapsulates the core tenets of TNA and accurately reflects
Markovian principles ---that is behind TNA--- where an event
(\emph{\textbf{occurrence}}) is contingent upon the preceding
event and tallied as transition matrices
\cite{Winne_Hadwin_Gress_2010}.
TNA captures these events as well as their transitions (\emph{\textbf{contingencies}})
and their typical patterns. However, TNA extends
\emph{\textbf{occurrences}} to include states (e.g., engagement,
tactics, or strategies), activities, utterances, and other categorical or discrete representations of learning or learners.

The mathematical underpinnings of TNA build on Markov processes. A Markov chain is a type of stochastic process that models a sequence of random variables, where the probability of transitioning to the next state (or contingency) depends only on the current state (or occurrence), not on any of the preceding states \cite{Sarkar_Moore_2005, Silver_Silva_2021,Spedicato_2017}. This defining feature is known as the Markov property, which can be expressed formally as follows: for all \( t \geq 1 \), \( \displaystyle
P(X_{t+1} = x \mid X_1 = x_1, \ldots, X_t = x_t) = P(X_{t+1} = x \mid X_t = x_t). \) A Markov model is a practical application of Markov chains, where we assume that the system under study can be represented by a Markov chain, and the states are directly observable. The primary goal of using a Markov model is to estimate the transition probabilities from observed data. When the state space \( S \) is finite, the transition probabilities can be organized into a matrix \( \mathbf{P} \), known as the transition matrix: \( \displaystyle
[\mathbf{P}]_{ij} = P(X_{t+1} = j \mid X_t = i), \quad i,j \in S, \quad t \geq 1.
\)
In this matrix, the element in the \( i \)th row and \( j \)th column represents the probability of transitioning from state \( i \) (i.e., the occurrence) to state \( j \) (i.e., the contingency). This transition matrix can also be interpreted as the adjacency matrix of a weighted directed graph \( G = (V, E, w) \), where \( V \) is the set of vertices, \( E \) is the set of directed edges, and \( w: E \rightarrow \mathbb{R}^{+} \) is a weight function assigning a weight to each edge, corresponding to the transition probabilities \cite{Sarkar_Moore_2005, Silver_Silva_2021}.

In the context of TNA, events and their transitions are inherently directional, making directed networks (or graphs) the appropriate mathematical structure to represent them. A directed network \( G \) is defined as a pair \( (V, E) \), where \( V = \{V_1, \ldots, V_n\} \) is the set of vertices (or nodes) and \( E \) is the set of edges. Each edge is a pair \( (V_i, V_j) \) where \( V_i, V_j \in V \), visually depicted as an arrow \( V_i \rightarrow V_j \). In this context, the nodes represent the states of a sequence, and the edges represent the transitions between these states \cite{Sarkar_Moore_2005, Silver_Silva_2021,Spedicato_2017}.

\subsection{Review of current learning analytics approaches}

Social Network Analysis (SNA) is a well-established method in learning
analytics, offering insights into how the structure of relationships and
interactions within learning environments impacts learning outcomes
\cite{Dawson_Tan_McWilliam_2011,Rienties2014-jy}.
SNA offers metrics such as centrality measures to analyze the role of
individuals in learning networks \cite{Saqr_López-Pernas_Conde_Hernández-García_2024}.
However, SNA's limitation lies in its inability to capture the
heterogeneity (clusters) or the temporality of interactions as it
predominantly analyzes the static structure of social ties without
addressing its dynamic evolution. More recently, Epistemic Network Analysis (ENA) has emerged as a method to capture the
connections between elements, such as concepts, skills, actions, and values \cite{Fernandez-Nieto2021-po,Kaliisa2023-xh}. ENA
operationalizes the theory of epistemic frames to model learning as a
network. In this framework, nodes represent coded interaction elements,
and edges represent their co-occurrence within a time window \cite{Shaffer_Collier_Ruis_2016}.
However, ENA primarily focuses on the
co-occurrence of elements and cannot model directed transitions between
learning states, limiting its ability to fully capture temporal learning
dynamics \cite{Tan_Swiecki_Ruis_Shaffer_2024}.
Ordered Network Analysis (ONA) can model directed networks,
allowing for the analysis of the order in which learning events occur \cite{Tan_Swiecki_Ruis_Shaffer_2024}.
While ONA addresses some gaps in ENA, both ENA and ONA have the same
shortcomings as SNA, i.e., lack of a method to capture the heterogeneity
of the network (e.g., clustering) or covariates that explain the
occurrence of certain patterns. Also, both ENA and ONA lack centrality
measures and do not offer a method for verifying the stability of the
edges offered by TNA or finding communities. Temporal Network Analysis can capture the evolution of interactions over time and
has proven valuable in understanding how the timing and duration of
interactions influence learning. Temporal metrics like dynamic
centrality have been shown to predict student performance
\cite{Saqr_Peeters_2022}.
However, temporal network analysis lacks a probabilistic framework for
modeling transitions between learning states as well as clustering and
covariate capabilities.

Process mining ---and especially SPM based on Markovian models--- shares some features with TNA. SPM has been commonly used to investigate and visualize the transitions between students' learning strategies \cite{Saint_Gaševic_Matcha_Uzir_Pardo_2020,Lopez-Pernas2021-di}, emotions in collaborative learning \cite{Tormanen2021-ds}, and time-management tactics \cite{Ahmad_Uzir2020-tt}. However, current implementations of SPM fall short of exploiting the full mathematical potential of graph modeling or offer a rigorous method for verifying the resulting models. TNA builds and extends the functionalities and possibilities of SPM beyond visualization by adding, e.g., network metrics, community finding, network pattern mining and model validity testing to provide deeper insights into the underlying
processes. While each method (SNA, ENA, ONA, SPM or temporal networks)
has significantly contributed to learning analytics \cite{López-Pernas_Misiejuk_Kaliisa_Conde-González_Saqr_2024}, they often operate
in isolation, focusing on specific aspects such as cognitive or social
dimensions or the temporal evolution of learning processes. TNA provides
an integrated framework that models transitions as directed, weighted
networks, incorporates covariates to explain why specific transitions
occur. Additionally, TNA introduces centrality measures to track the
significance of particular learning events and detect clusters of
learners based on their transition patterns. As such, TNA adds to the
methods in learning analytics and is particularly suited for analyzing dynamic
or event-based processes. Below we offer an example case study in a
collaborative environment. Given the limitations of the space, a full
detailed description of the empirical findings is beyond this paper; therefore, we will emphasize the methods and their applications.
\subsection{Case study: TNA to reveal the dynamics of small-group collaboration}

Previous research has shown that productive small-group collaboration
requires students\textquotesingle{} ability to engage in the regulation
of learning, that is, to plan, monitor, and control cognitive,
motivational, emotional, and social aspects of group work
\cite{Hadwin_Bakhtiar_Miller_2018}.
In a group context, this responsibility is a joint one: group members
need to engage in socially shared regulation of learning (SSRL), where
they jointly negotiate and build on each other's contributions to set
shared goals, enact strategies, monitor their progress, and evaluate if
strategic changes are needed to reach the set goals
\cite{Hadwin_Bakhtiar_Miller_2018}.
Moreover, if individual group members face challenges in their learning
process, other group members can enact co-regulation to support them
\cite{Hadwin_Bakhtiar_Miller_2018}.
Together, SSRL and co-regulation can be conceptualized as occurring at
the group level because the regulatory actions need to be verbalized in
the group's social interactions and, therefore, are observable from the
group\textquotesingle s verbal or written interactions
\cite{Törmänen_Järvenoja_Saqr_Malmberg_Järvelä_2023}.
Increasing theoretical understanding of group-level regulation indicates
that productive regulation involves the temporal interplay of various
social (e.g., participation, see
\cite{Vuorenmaa_Järvelä_Dindar_Järvenoja_2023}),
cognitive (e.g., knowledge construction, see
\cite{Zabolotna_Malmberg_Järvenoja_2023}),
and emotional (e.g., emotional states, see
\cite{Törmänen_Järvenoja_Saqr_Malmberg_Järvelä_2023})
aspects of group learning. However, the majority of previous studies
have investigated these processes as separate, being unable to capture
the complex interplay between these multiple components. Also, research
addressing the temporality of these processes to, for example, identify
the productive patterns of groups' social interactions, is still scarce.
TNA could provide a means to capture the temporal changes in group
interactions, identify patterns that lead to successful collaboration
and derive insights into the different
components and layers of SSRL and co-regulation. Thus, this case study addresses the following research questions: 

RQ1: \emph{What can TNA reveal about students'
regulation approach in project-based learning in terms of the overall
dynamics (I), patterns of regulation, and the central regulatory
processes (II), the covariates that explain the patterns (III), how
high achievers differ from low achievers (IV), and what roles students
play in the collaboration (V)?} RQ2: \emph{How stable is the model estimated by TNA?}

\section{Methods}

The data used to demonstrate TNA was gathered from three offerings of a
master-level course on computer science. In these course offerings, students
had to collaborate together online using chats to develop and
deliver a project. The project involves collecting data, analyzing it as
well as presenting the results by the end of the course. The number of
students was 191 interacting in 24 small groups. There were 10,543
interactions in total, with an average of 439.3 interactions per group
(SD=282.9). There were 8.1 students per group on average (SD=0.9),
with an average of 55.2 messages per student (SD=68.4). 

The students\textquotesingle{} online chat discussions were collected and coded with the coding scheme informed by constructs from SSRL and coregulation perspectives covering different elements related to \textbf{\emph{task enactment} }(i.e., students enact various tactics and strategies to reach their goal \cite{Winne_Hadwin_1998}), \textbf{\emph{socio-emotional interactions}} (i.e., students’ express emotions or take other actions contributing to socio-emotional aspects of group work (e.g., group formation and dynamics, social relationships, a sense of community \cite{Törmänen_Järvenoja_Saqr_Malmberg_Järvelä_2023}), and \textbf{\emph{group-level regulation}} (i.e., students’ express the observation of an obstacle or a challenge in their learning process, make a regulatory initiation to activate control, which can lead to strategic change in the groups' action \cite{Hadwin_Bakhtiar_Miller_2018}) (see Table \ref{table:codingscheme}). After 10\% of interactions coded by two researchers achieved at least substantial reliability ($\kappa$\textgreater0.61), the rest of the data was split and coded by individual researchers.

\subsection{Implementation of the method}

Students' coded interactions were grouped into sessions where posts that
received replies within 20 minutes (corresponding to the 90\textsuperscript{th} quantile
of gaps of activity) were grouped together
\cite{Jovanovic2017-uk}.
The time-ordered sequence of each group coded interactions was used for the estimation of
the Markov model, and in all subsequent analyses. We estimated the
Markov model using the R packages \texttt{tna} \cite{Anonymous_2024} and \texttt{seqHMM} \cite{Helske_Helske_2019,Spedicato_2017} where the
input was the sequence object created in the previous step. In a simple
Markov model, initial state probabilities, and transition probabilities
are estimated directly from the data as relative frequencies of initial
state memberships and transitions between states. To capture the different transition patterns, we used clustering. Clusters refer to subgroups within the data that exhibit distinct transition configurations between states as reflections of the underlying process.

\begin{small}
\begin{longtable}[t]
{p{1.4cm}p{3cm}p{1.9cm}p{4.5cm}rrr}

\caption{Coding scheme for chat discussions.}
\vspace{-3mm}
\label{table:codingscheme} \\ \hline
\textbf{Elements}  & \textbf{Description} & \textbf{Code} & \textbf{Clarification} & \textbf{N} & \textbf{M} & \textbf{SD} \\ \hline
\textbf{Task \newline enactment} \newline  & \multirow{3}{3cm}{Students enact various tactics and strategies to reach their goal \cite{Winne_Hadwin_1998}.} & \textbf{\textcolor{plan}{Planning}} & Students coordinate different aspects of the project by e.g., dividing tasks and scheduling meetings. & 2402 & 100.08 & 63.37 \\ \cline{3-7}
 & & \textbf{\textcolor{explore}{Exploring}} & Students are sharing information and ideas to construct a shared understanding of a problem. & 1704 & 71.00 & 70.30 \\ \cline{3-7}
 & & \textbf{\textcolor{synthesis}{Synthesis}} & Students synthesize the available information to reach a joint understanding or solution. & 313 & 13.61 & 10.95 \\ \hline
\textbf{Socio-emotional interaction} & \multirow{3}{3cm}{Students’ express emotions or take other actions contributing to socio-emotional aspects of group work (e.g., group formation and dynamics, social relationships, a sense of community) \cite{Törmänen_Järvenoja_Saqr_Malmberg_Järvelä_2023}.} & \textbf{\textcolor{emotion}{Emotion}} & Students express their emotions either textually or by using emojis. & 1162 & 48.42 & 34.07 \\
\cline{3-7}
 & & \textbf{\textcolor{understand}{Understanding}} & Students show understanding by acknowledging and respecting each others’ views, ideas, and contributions. & 2627 & 109.46 & 79.73 \\ \cline{3-7}
 & & \textbf{\textcolor{cohesion}{Cohesion}}  &  Students foster group formation and sense of community by encouraging joint participation and cohesive group interactions. & 694 & 28.92 & 21.77 \\ \hline
\textbf{Group \newline regulation} \newline & \multirow{3}{3cm}{Students express the observation of an obstacle or a challenge in their learning process, make a regulatory initiation to activate control, which can lead to strategic change in the groups' action \cite{Hadwin_Bakhtiar_Miller_2018}.} & \textbf{\textcolor{monitor}{Monitoring}}  & Students monitor their task progress and identify issues or problems that need to be addressed. & 623 & 25.96 & 17.36 \\ \cline{3-7}
 & & \textbf{\textcolor{coregulate}{Co-regulation}} & Students control the group’s cognitive, motivational, emotional, or social learning activities to ensure progress towards the shared goal. & 783 & 32.62 & 25.45 \\ \cline{3-7}
 &  & \textbf{\textcolor{adapt}{Adaptation}} &  Students make a strategic change and apply new ideas, tactics, or strategies to solve problems and resolve issues at hand. & 235 & 11.19 & 12.32 \\ \hline 
\end{longtable}

\end{small}
%\end{landscape}

Using the \texttt{seqHMM} package, we
estimated several clustering models with varying numbers of clusters (2-8) and the
model with lowest BIC value was selected (BIC:2 clusters=36095.89, BIC:3 clusters=37209.43, BIC:4 clusters=35975.73, BIC:5 clusters=37288.99, BIC:6 clusters=38999.07, BIC:7 clusters=failed, BIC:8 clusters=failed). We fitted the models including three covariates: average group grades (to account for achievement differences), course (to account for course variability), and
number of group members (given that the number of collaborators may affect the group dynamics and the division of work), using the Expectation-Maximization
algorithm. The model was optimized with 500 restarts, employing both
global and local optimization steps. Out of 500 restarts, the best model
was reached 486 times, confirming the robustness of the optimization and
consistency in reaching the global optimum. While several centrality
measures may be computed, for the sake of demonstrating, we computed
\emph{in-strength centrality} (the sum of incoming transition weights),
which indicates the importance of a key regulation process for students.

We also computed \emph{betweenness centrality}, measuring the frequency
with which a node lies on the shortest paths between other nodes and
thus, mediating ---or bridging--- the transitions between different SRL
processes. The betweenness centrality was computed using the random walk betweenness centrality measure because random walk processes (such as Markov chains) are more aligned with our context of probabilistic networks and stochastic processes \cite{DePaolis2022-bd}.
We also extracted patterns from our model using the \texttt{tna} package: \emph{dyads} (mutual
transitions with a threshold of 0.1 in either direction), and \emph{cliques}
(triplets of three nodes where transitions occur above a threshold of
0.05 in all directions) ---please note that these thresholds were set for
demonstration based on our contextual case given the lack of guidance on optimal threshold. 
To mine the network communities, we chose the Spin-glass community-finding algorithm given its robustness to noise, ability to detect small communities, and suitability for directed weighted networks \cite{PhysRevE}.

\section{Results}

\subsection{{TNA plots and network measures: Group regulation dynamics}}

\begin{figure*}[!ht]
 \centering
 \includegraphics[width=0.9\linewidth]{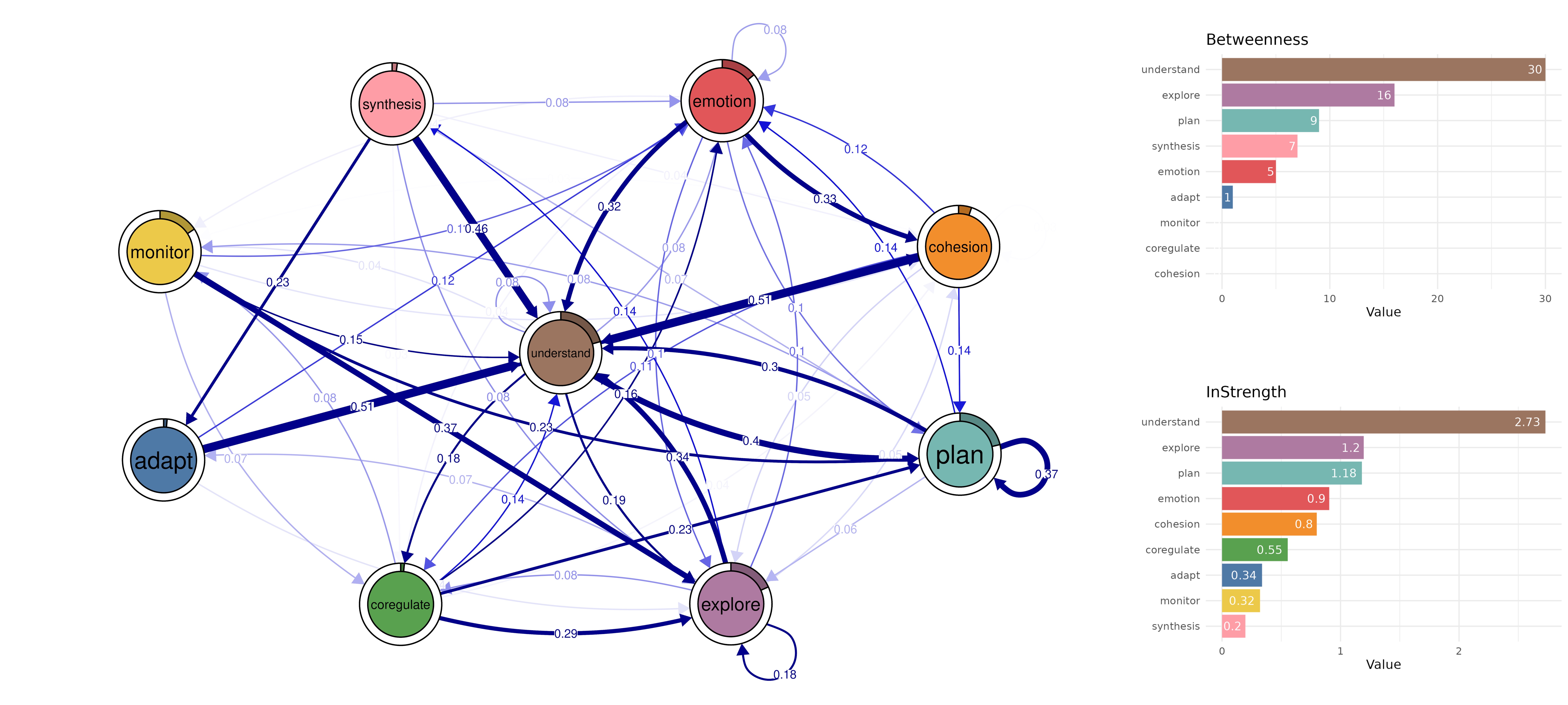}
  \vspace{-5mm}
 \caption{\textbf{Left}: A TNA plot of group collaborative interactions in a
project; \textbf{Right top}: A bar chart representing the betweenness centrality
of each code; \textbf{Right bottom}: A bar chart showing the in-strength
centralities of each code.}
\label{fig:centralities}
\end{figure*}

A TNA plot in Figure \ref{fig:centralities} shows how students regulated their project-based interactions (RQ1-I). \emph{Nodes} represent different regulatory processes in our case study. Each node has a pie which represents the initial probability (the initial start before any transition). \emph{Edges}
represent the transition from one process to another through arrows that indicate the direction of the transition and the thickness and the
numbers show the probabilities of these transitions. The TNA plot captures students' collaborative interactions during their project work.
The transition process shows the strong connection between \textbf{task
enactment} and \textbf{socio-emotional interaction}. \textcolor{understand}{\emph{Understanding}} had the
highest in-strength and betweenness values, making it the most central in students\textquotesingle{} collaborative interactions. It bridges other transitions and receives many transitions itself. For example, students frequently showed \textit{\textcolor{understand}{understanding}} after fostering group
\textit{\textcolor{cohesion}{cohesion}} or reaching a joint solution through \textit{\textcolor{synthesis}{synthesis}}.
Following \textit{\textcolor{understand}{understanding}}, \textbf{task enactment} nodes (\textit{\textcolor{explore}{exploring}},
\textit{\textcolor{plan}{planning}}, \textit{\textcolor{synthesis}{synthesis}}) received the most transitions, with \textit{\textcolor{explore}{exploring}}
being particularly central and connected to \textbf{socio-emotional
interactions}. After sharing information through \textit{\textcolor{explore}{exploring}},
students were likely to acknowledge each other's views by showing
\textit{\textcolor{understand}{understanding}}. In contrast, \textbf{group regulation} processes, such as
\textit{\textcolor{monitor}{monitoring}} and \textit{\textcolor{adapt}{adaptation}}, had relatively few transitions.
However, \textit{\textcolor{monitor}{monitoring}} and \textit{\textcolor{coregulate}{co-regulation} }still prompted
information sharing: \textit{\textcolor{explore}{exploring}} followed \textit{\textcolor{monitor}{monitoring}} and
\textit{\textcolor{coregulate}{co-regulation}} with high transition probability. In general, the transition process shows a strong emphasis on \textbf{task enactment} and \textbf{socio-emotional interaction}, with a limited role for metacognitive or deep cognitive processes, such as small-scale adaptation. 

%\vspace{-8mm}

\subsection{Patterns: Dyads, cliques, communities and clusters}

According to Winne's framework ---that underpins our study--- patterns are repeated regular arrangements of transitions that shape students' choices and behavior. ``Researchers assume mental operations generate behavior, and those cognitive and metacognitive events are what theories seek to account for'' \cite[p.~272]{Winne_2010}. When patterns are regular, repeated, and occur frequently, they are indications of tactics and SRL typical behavior \cite{Chen_Poquet_2020,Kwon_Liu_Johnson_2014,Poquet_Dawson_Dowell_2017}. TNA allows the identification of several types of patterns: dyads, cliques, communities, and clusters (RQ1-II).

\begin{figure*}[!ht]
 \centering
 \includegraphics[width=.95\linewidth]{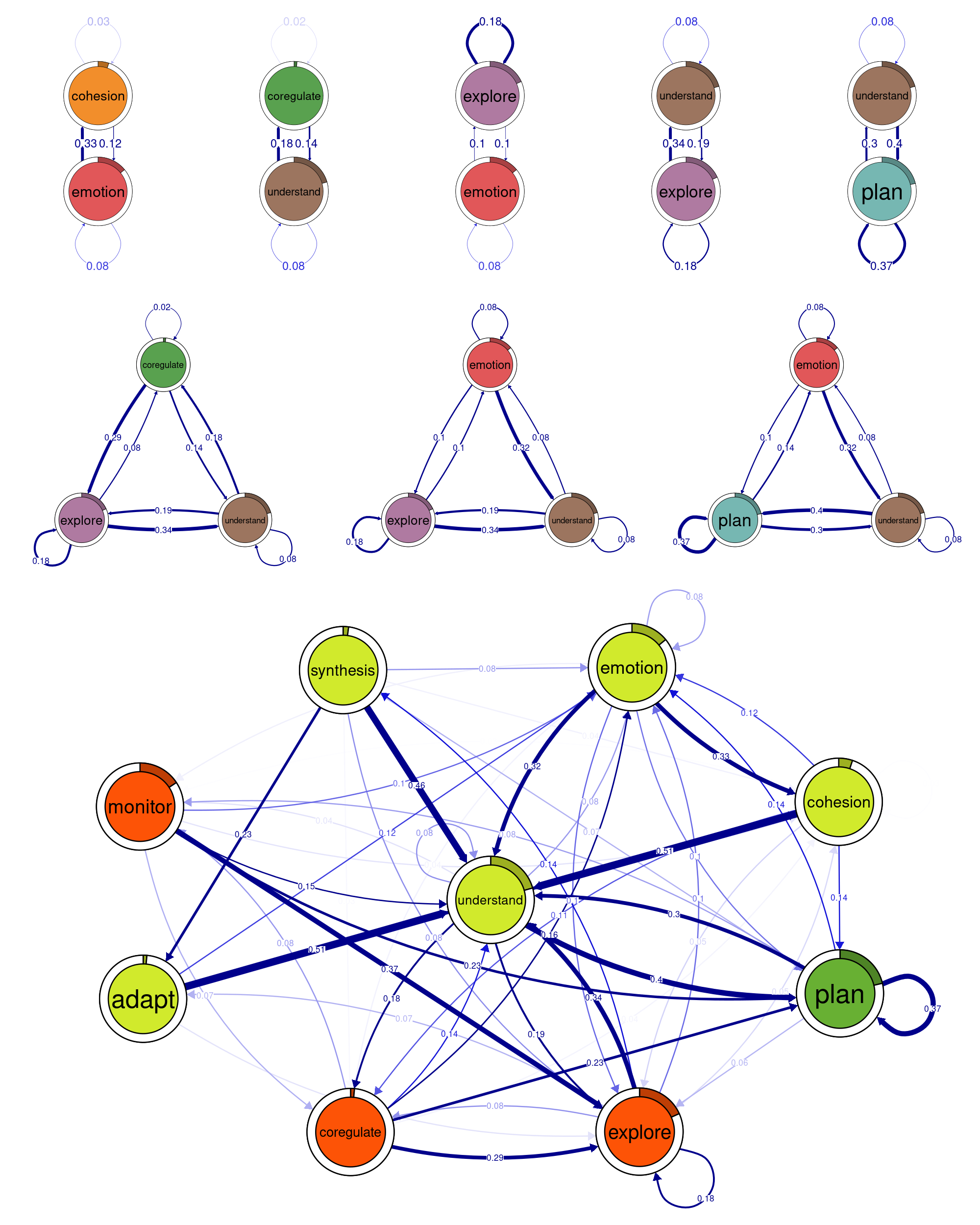}
  \vspace{-5mm}
 \caption{\textit{Dyads} (top), \textit{triads} (middle), and \textit{communities} (bottom)}
 \label{fig:dyads}
\end{figure*}

\emph{\textbf{Dyads}} are transitions between a pair of nodes, when they
are mutual and have high probabilities (we use the threshold of 0.1
here), they represent likely interdependent processes. In our case
study, we see several \textit{dyads} (see Figure \ref{fig:dyads} - top).
\emph{\textcolor{cohesion}{Cohesion}}-\emph{\textcolor{emotion} {emotion}} (0.33-0.12) and \emph{\textcolor{explore}{exploring}}-\emph{\textcolor{emotion}{emotion}} (0.1-0.1) reveal the role of emotions in
strengthening a sense of community and information sharing in groups.
All other strongly connected \textit{dyads} involve task enactment and group
regulation nodes, such as \emph{\textcolor{coregulate}{coregulate}}-\emph{\textcolor{understand}{understanding}}
(0.18-0.14), or \textbf{task enactment} and \textbf{socio-emotional interaction}
nodes, such as \emph{\textcolor{explore}{exploring}}-\emph{\textcolor{understand}{understanding}} (0.34-0.19) and
\emph{\textcolor{understand}{understanding}}-\emph{\textcolor{plan}{planning}} (0.3-0.4), demonstrating how \textbf{task
enactment} is the main focus of group interaction in a collaborative
task. 

\emph{\textbf{Triads}} (often referred to as \textit{cliques} when all nodes are
connected to each other forming a complete sub-graph like in our case here) are constellations of three connected nodes and reflect ``recurring, significant patterns of
interconnections in the network'' \cite[p.~500]{Chen_Zhu_Shui_2022}. Identifying cliques can help understand which actions or states characterize the dynamics between \textbf{task enactment}, \textbf{socio-emotional
interactions}, and \textbf{group regulation }\cite{Chen_Zhu_Shui_2022}. In our TNA network, three cliques
existed above the weight of 0.05 in both directions (see Figure \ref{fig:dyads} -
middle). The first clique (\emph{\textcolor{coregulate}{co-regulation}}-\emph{\textcolor{explore}{exploring}-\textcolor{understand}{understanding}})
shows the interplay between \textbf{task enactment}, \textbf{socio-emotional interactions}
and \textbf{group-level regulation}. The second clique (\textcolor{emotion}{\emph{emotion}}-\emph{\textcolor{explore}{exploring}}-\emph{\textcolor{understand}{understanding}}) highlights how emotions are linked to \textbf{task enactment} and showing understanding in the group, and the third clique (\emph{\textcolor{emotion}{emotion}}-\emph{\textcolor{plan}{planning}}-\emph{\textcolor{understand}{understanding}}) shows \textbf{socio-emotional interactions} are intertwined with \textbf{task enactment}, particularly \emph{\textcolor{plan}{planning}}.
Notably, \emph{\textcolor{understand}{understanding}} is induced in all the triangles.

\emph{\textbf{Communities}} are another type of pattern that groups edges according to their stronger transition within their communities (see Figure \ref{fig:dyads} - bottom). In our case study, three communities emerged. The \textbf{largest community} denoted by the \textcolor{yellowcom}{light yellow} color includes five nodes and shows the importance of \textbf{socio-emotional interactions} (\textcolor{understand}{\emph{understanding}}, \textcolor{emotion}{\emph{emotion}}, \textcolor{cohesion}{\emph{cohesion}}) for students synthesizing available information to reach joint understanding (\textcolor{synthesis}{\emph{synthesis}}) and adapting their strategies to improve problem-solving (\textcolor{adapt}{\emph{adaptation}}). The \textbf{second community} (\textcolor{orangecom}{dark orange}) shows a regulation pattern including \textcolor{coregulate}{\emph{co-regulation}} and \textcolor{monitor}{\emph{monitoring}} that aids students in sharing information and ideas (\textcolor{explore}{\emph{exploring}}). Finally, \textcolor{plan}{\emph{planning}} is in a community of its own (the \textbf{third community}, in \textcolor{greencom}{green}), revealing the temporal order where planning was more likely to be at its own time
scale.
\begin{figure*}[!ht]
 \centering
 \includegraphics[width=.9\linewidth]{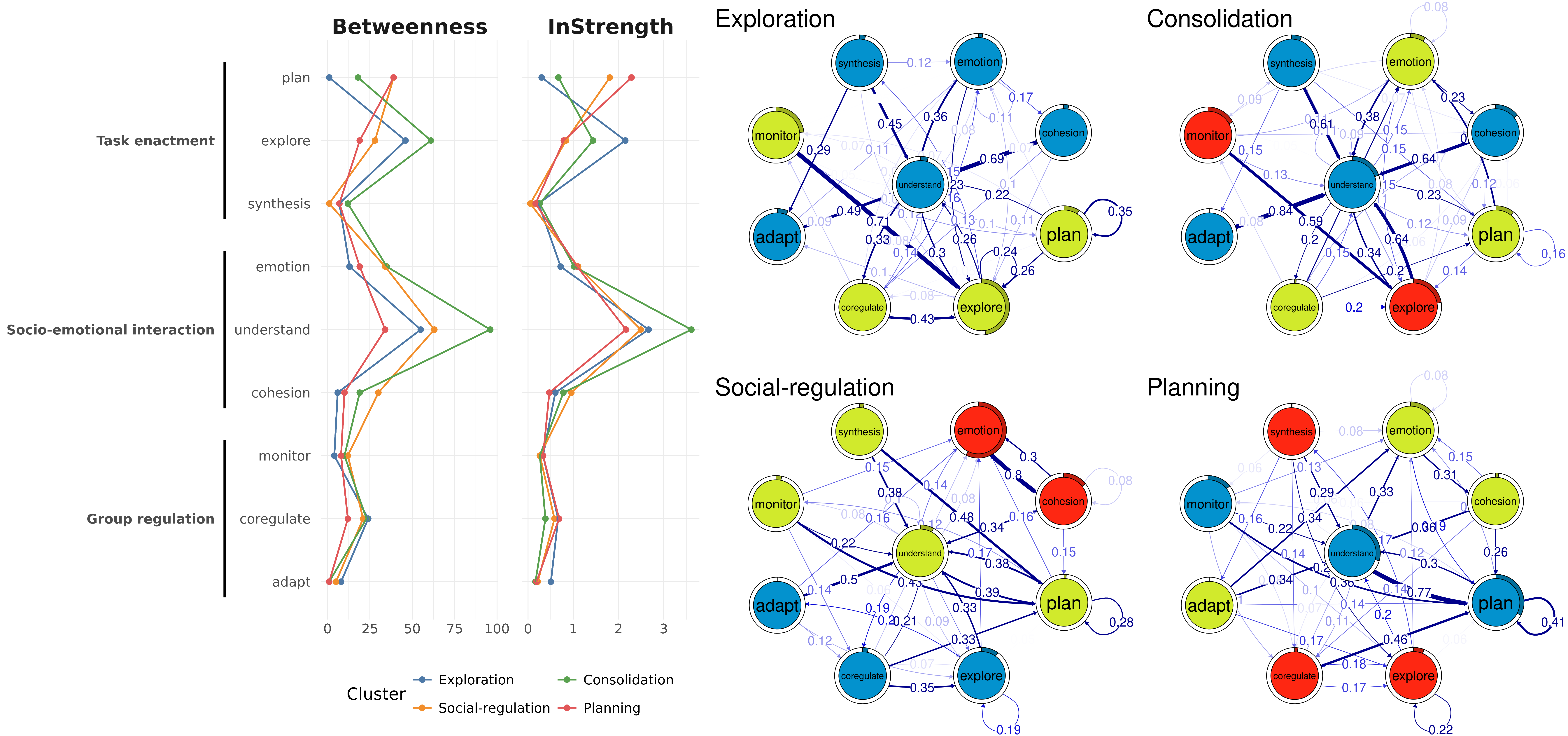}
  \vspace{-2mm}
 \caption{ \textbf{Left}: centrality measures of each cluster. \textbf{Right}: Transition
network of each cluster}
\label{fig:clustercentralities}
\end{figure*}

\emph{\textbf{Clusters}} are groups of sessions that follow a similar pattern (periods), or what is known as \textit{tactics} in learning analytics research \cite{Matcha_Gašević_Ahmad_Uzir_Jovanović_Pardo_Maldonado-Mahauad_Pérez-Sanagustín_2019,Saqr_López-Pernas_Jovanović_Gašević_2023, Lopez-Pernas2021-di}. Of course, clustering is not a new technique, but it has been commonly ---and, in fact, overwhelmingly--- used at the sequence level. Here, we extend it with other functionality like community finding and centrality to better characterize these clusters. In our dataset, we identified four clusters that represent different
types of sessions during student project work (see Figure \ref{fig:clustercentralities}).

\begin{itemize}
  \item The \textbf{exploration cluster} is focused on task enactment, particularly \textcolor{explore}{\emph{exploring}} and The \textbf{socio-emotional interaction} (\textcolor{understand}{\emph{understanding}}) with high in-strength centrality. \textcolor{coregulate}{\emph{Co-regulation}} is prominent with relatively high betweenness and in-strength centrality, while other regulation processes have low interactions. Within this cluster, we identified two communities: one that involves \textcolor{plan}{\emph{planning}}, \textcolor{coregulate}{\emph{co-regulation}}, \textcolor{explore}{\emph{exploring}} and \textcolor{monitor}{\emph{monitoring}} and the other community including \textcolor{synthesis}{\emph{synthesis}}, \textbf{socio-emotional interactions}, and \textcolor{adapt}{\emph{adaptation}}.
  
  \item The \textbf{consolidation cluster} has prominent socio-emotional interactions to facilitate \textcolor{explore}{\emph{exploring}}, \textcolor{coregulate}{\emph{co-regulation}} and \textcolor{synthesis}{\emph{synthesis}}, indicating an emphasis on consolidating the project plans as well as building stronger relationships among collaborators. Three communities were identified in this community: one built around \textcolor{understand}{\emph{understanding}} to foster \textcolor{cohesion}{\emph{cohesion}}, making strategic changes through \textcolor{adapt}{\emph{adaptation}}, and reaching \textcolor{synthesis}{\emph{synthesis}}, second that includes \textcolor{explore}{\emph{exploring}} and \textcolor{monitor}{\emph{monitoring}}, and third where \textcolor{emotion}{\emph{emotions}} are connected to \textcolor{coregulate}{\emph{co-regulation}} and coordination of the project work (\textcolor{plan}{\emph{planning}}).
  
  \item The \textbf{socio-emotional cluster} starts most of the time in either \textcolor{cohesion}{\emph{cohesion}} or \textcolor{emotion}{\emph{emotion}} and has strong bidirectional connections between both processes. We see a community of group \textcolor{cohesion}{\emph{cohesion}} and \textcolor{emotion}{\emph{emotions}}, another including \textcolor{understand}{\emph{understanding}} together with task enactment processes of \textcolor{plan}{\emph{planning}} and \textcolor{synthesis}{\emph{synthesis}}, and connected with \textcolor{monitor}{\emph{monitoring}}. The third community involves \textcolor{coregulate}{\emph{co-regulation}} and \textcolor{adapt}{\emph{adaptation}} connected with \textcolor{explore}{\emph{exploring}}.
  
  \item The \textbf{planning cluster} showcases student interactions starting frequently with \textcolor{plan}{\emph{planning}} and considerable transitions to \textcolor{plan}{\emph{planning}} (loop) as well as to \textcolor{understand}{\emph{understanding}} and \textcolor{monitor}{\emph{monitoring}}, forming a community. \textcolor{plan}{\emph{Planning}} also has the highest in-strength and betweenness centrality in this cluster. The second community includes \textcolor{emotion}{\emph{emotion}}, \textcolor{cohesion}{\emph{cohesion}} and \textcolor{adapt}{\emph{adaptation}}, and the third one \textcolor{explore}{\emph{exploring}}, \textcolor{synthesis}{\emph{synthesis}}, and \textcolor{coregulate}{\emph{co-regulation}}.
\end{itemize}
\subsection{Subtraction plots: High achievers vs. Low achievers}

\begin{figure}[!ht]
 \centering
 \includegraphics[width=.99\linewidth]{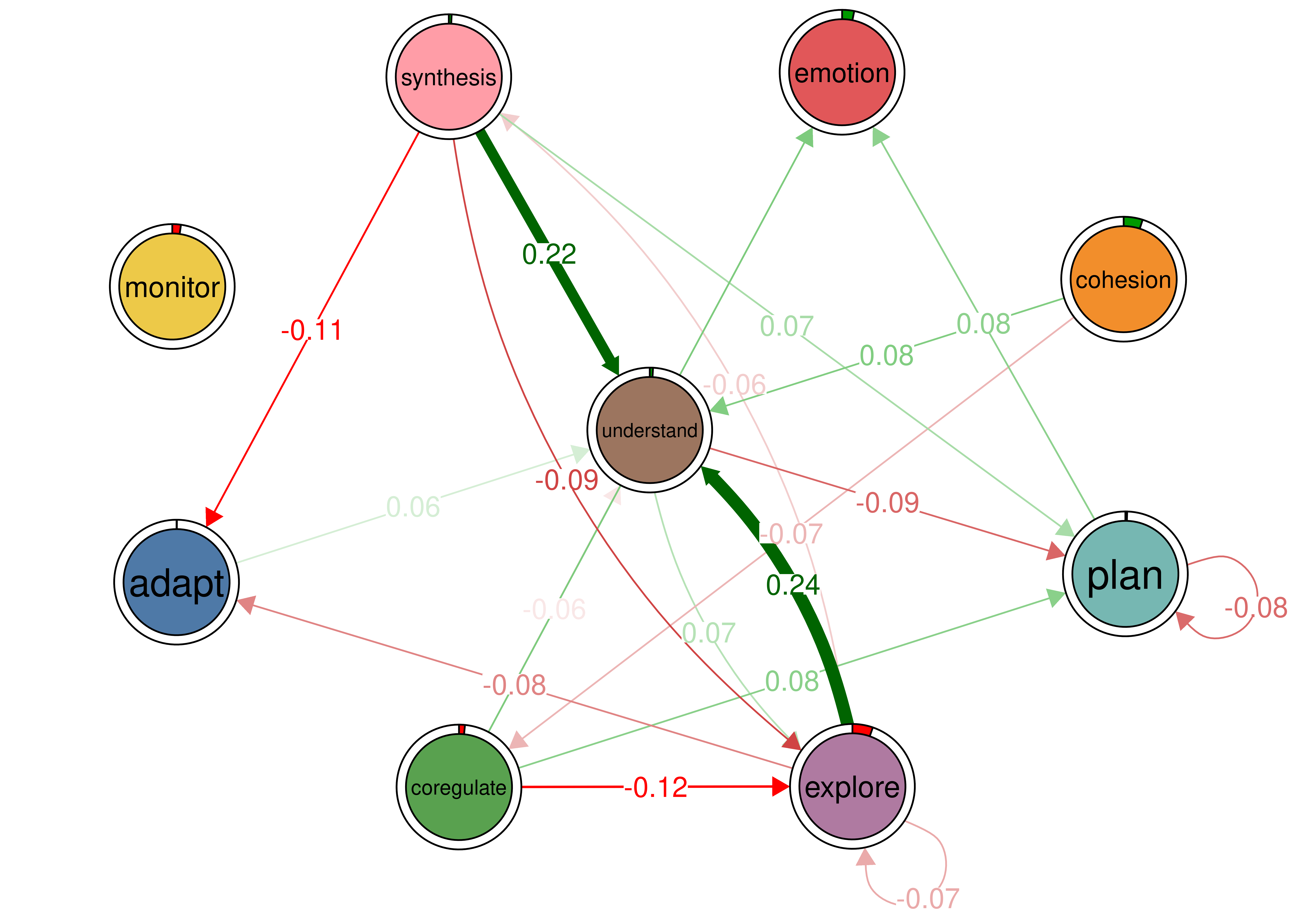}
 \vspace{-5mm}
 \caption{Subtraction TNA network comparing high- and low-performing
students that shows transition probabilities of high-performing
students.}
\label{fig:subtraction}
\end{figure}

Similarly to ENA, TNA enables the creation of subtraction plots to
compare transition processes between different groups (RQ1-IV). In Fig. \ref{fig:subtraction}, we
see that \textbf{high-performing students} were more likely to transition from
synthesizing information to building shared understanding
(\emph{\textcolor{synthesis}{synthesis}-\textcolor{understand}{understanding}}, t.p.=+0.22), while building shared
understanding transitioned to further explorations
(\emph{\textcolor{understand}{understanding}-\textcolor{explore}{exploring}}, t.p.=+0.24) than low-performing students.
At the same time, \textbf{high-performing students} were less likely to
transition from synthesizing information to changing their strategy to
problem-solving (\emph{\textcolor{synthesis}{synthesis}-\textcolor{adapt}{adaptation}}, t.p.=-0.11) or to jump to
further exploration (\emph{\textcolor{synthesis}{synthesis}-\textcolor{explore}{exploring}}, t.p.=-0.09) than
low-performing students. In comparison to \textbf{low-performing students},
\emph{\textcolor{coregulate}{co-regulation}} was less likely to be followed by \emph{\textcolor{explore}{exploring}}
(t.p.=0.12) and \emph{\textcolor{understand}{understanding}} was less likely to lead to
\emph{\textcolor{plan}{planning}} (t.p.=-0.09) in high-performing students.

\subsection{Idiographic plots: Students' roles}
\begin{figure*}[!hbt]
 \centering
 \includegraphics[width=.99\linewidth]{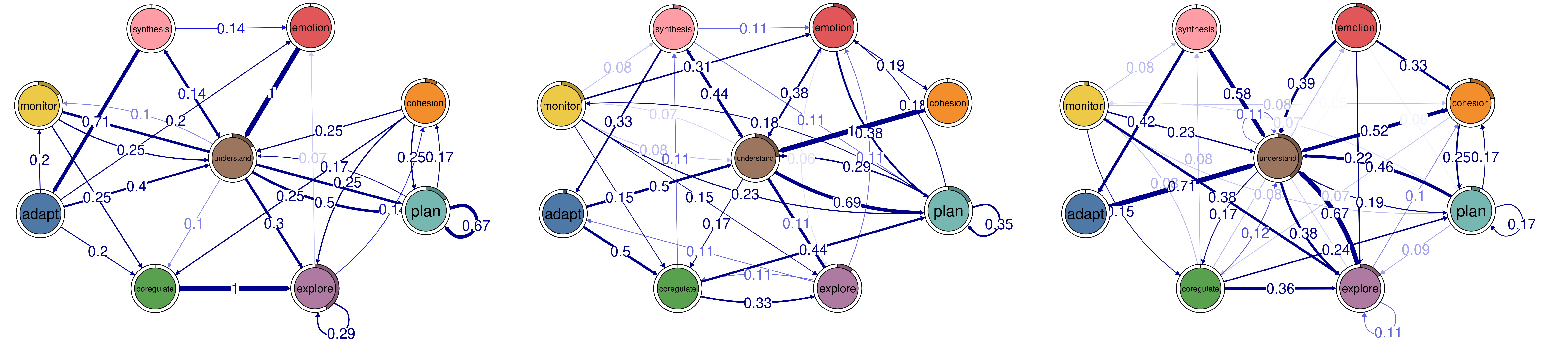}
 \vspace{-5mm}
 \caption{Idiographic plots of three different students: \textit{planner} (left), \textit{regulator} (middle), \textit{socializer} (right).}
\label{fig:idiographicplots}
\end{figure*}

TNA can facilitate idiographic analysis, i.e., it can build a full model
of a single student\textquotesingle s interactions ---given that enough 
data is available (RQ1-V). In Fig. \ref{fig:idiographicplots}, the first student can be seen to start
with \textit{\textcolor{plan}{planning}} and continue with \textit{\textcolor{plan}{planning}} which highlights their role as
a \emph{\textbf{planner}}. The second student starts more in \textit{\textcolor{monitor}{monitoring}}, and has
prominent transitions to \textit{\textcolor{coregulate}{co-regulation}}, \textit{\textcolor{adapt}{adaptation}} and \textit{\textcolor{plan}{planning}},
highlighting their role as \emph{\textbf{regulator}}. The third student starts
frequently and has eminent transitions to and from \textit{\textcolor{cohesion}{cohesion}} and \textit{\textcolor{emotion}{emotion}}
which highlight their role as a \emph{\textbf{socializer}}.

\vspace{-3mm}
\subsection{Covariate analysis: Explaining clusters}

Covariate analysis can explain the emergence of a certain cluster (RQ1-III). For instance, compared to the \textcolor{explorationcluster}{\textbf{exploration}} cluster (the baseline for comparison), students
with higher grades, and larger numbers of students were more likely to
engage in \textcolor{socialcluster}{\textbf{social regulation}} activities (see Table \ref{table:covariateanalysis}). Also, the \textcolor{consolidationcluster}{\textbf{consolidation}} cluster was more likely to emerge with a higher number of students in the group. The \textcolor{planningcluster}{\textbf{planning}} cluster was more likely to emerge in groups with a higher number of students indicating
the need for coordination.
%\vspace{-8mm}

\begin{table}[htbp]
\scriptsize
\centering
\caption{Results of the covariate analysis.}
\label{table:covariateanalysis}
\begin{xtabular}{llrrrrl}
\hline
\textbf{Cluster} & \textbf{Variable} & \textbf{Estimate} & \textbf{CI} & \textbf{t} & \textbf{p} & \\ \hline

\textcolor{consolidationcluster}{\textbf{Consolidation}} & Course 1 & -5.55 & [-9.59;-1.5] & -2.69 & 0.01 & \hspace{-3mm}** \\
& Course 2           & -2.72 & [-6.96;1.51]   & -1.26 & 0.21 & \\
& Course 3           & -6.52 & [-10.25;-2.78] & -3.42 & 0.00 & \hspace{-3mm}*** \\
& No. of students    & 0.40  & [0.09;0.71]    & 2.54  & 0.01 & \hspace{-3mm}** \\
& Mean group grade   & 0.05  & [-0.64;0.74]   & 0.14  & 0.89 & \\
\hline
\textcolor{socialcluster}{\textbf{Social-regulation}} & Course 1 & -11.15 & [-16.45;-5.84] & -4.12 & 0.00 & \hspace{-3mm}*** \\
& Course 2               & -12.01   & [-17.65;-6.37] &  -4.17 & 0.00 & \hspace{-3mm}*** \\
& Course 3               & -11.24   & [-16.04;-6.43] & -4.59 & 0.00 & \hspace{-3mm}*** \\
& No. of students        & 0.66     & [0.29;1.02] & 3.55 & 0.00 & \hspace{-3mm}*** \\
& Mean group grade       & 1.20     & [0.43;1.96] & 3.06 & 0.00 & \hspace{-3mm}*** \\
\hline
\textcolor{planningcluster}{\textbf{Planning}} & Course 1 & -1.87 & [-4.92;1.18] & -1.20 & 0.23 & \\
& Course 2              & -2.85     & [-6.08;0.37] & -1.73 & 0.08 & \\
& Course 3              & -2.37     & [-5.11;0.38] & -1.69 & 0.09 & \\
& No. of students       & 0.29      & [0.07;0.51] & 2.60 & 0.01 & \hspace{-3mm}** \\
& Mean group grade      & 0.23      & [-0.21;0.66] & 1.02 & 0.31 & \\  \hline 

\end{xtabular}
\end{table}

\subsection{Validation of TNA plots: Model stability}
\begin{figure*}[!ht]
 \centering
 \includegraphics[width=.99\linewidth]{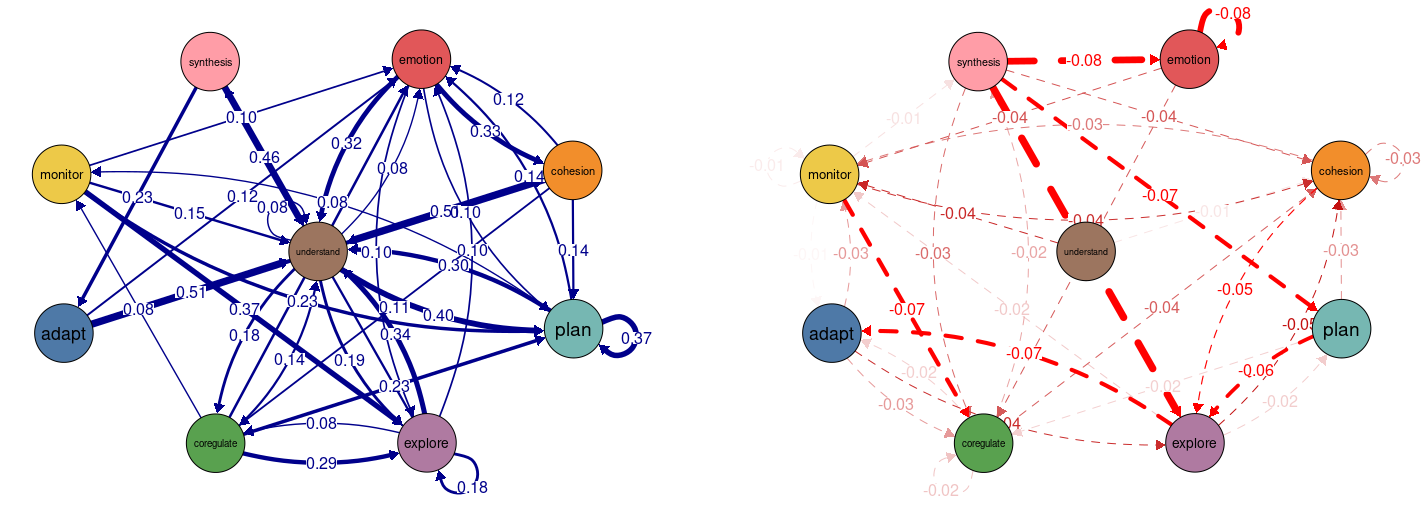}
 \caption{\textbf{Left}: TNA plot after bootstrapping. \textbf{Right}: TNA visualization of the
edges that were deleted by bootstrapping indicating that these
dropped edges may not replicate in future iterations.}
\label{fig:bootstrap}
\end{figure*}

Validation of process models or network edges are hardly visible in the
learning analytics literature and therefore, most existing SPM models are descriptive.
Descriptive models are useful in showing, for instance, how students
approached their project and how teachers can intervene in them as we based
our analysis. However, in many instances, we need to know how stable our
models are and how they may replicate in future iterations (RQ2). TNA uses
bootstrap methods to find significant edges, given its ability to
estimate uncertainty, facilitate statistical inference, and extinguish
weak edges. In bootstrapping, the data is repeatedly resampled and a TNA
model is estimated, edges that consistently appear across most bootstrap
samples are likely to be more stable and significant. Bootstrap methods
allow for the calculation of confidence intervals or p-values for edge
weight, helping to exclude spurious edges and estimate the robustness of
each in the network \cite{Mooney1993-az}. As a non-parametric method, bootstrap does not
require strong distributional assumptions. Given that transition
networks are fully connected and lack sparsity, bootstrapping offers a
credible method for retrieving the key backbone of the network,
eliminating the negligible edges, and making the networks easier to
interpret and understand. We applied bootstrapping to the current
network which eliminated 42 edges, most of which were small (mean
t.p.=0.03) (see Figure \ref{fig:bootstrap}). However, several edges were eliminated namely edges outgoing from \emph{\textcolor{synthesis}{synthesis}}, incoming to \textcolor{cohesion}{\emph{cohesion}}, \emph{\textcolor{explore}{exploring}} and \emph{\textcolor{coregulate}{co-regulation}}. Although these edges had
relatively high probabilities. While these edges exist in the
descriptive model, they are unstable. In other words, they are less
likely to replicate in other samples.

\section{Discussion}

The 21\textsuperscript{st} century marked a shift in the adoption of digital technologies driven by the
abundance of data and the development of new research methodologies. These advancements spurred
the emergence of learning analytics where several techniques were
adapted, applied, and expanded to explore the dynamics of learning
processes. Techniques, such as SNA, ENA, and ONA have all been used to capture the learning process. Each of
these methods was used to offer unique insights. In this paper, we
presented TNA as a novel framework for the analysis of event data as a
temporal process. TNA builds  ---and expands--- on the strong foundations
of the existing models.

In the grand scheme of things, the use of networks to model or represent the
stochastic transition process or Markov models are neither entirely new
nor uncommon. In fact, several seminal papers on Markov models
have used networks explicitly to represent and visualize the transition
process and the literature is awash with empirical examples
\cite{Gales_Young_2008,Kim_Lee_Xue_Niu_2018,Sarkar_Moore_2005,Silver_Silva_2021,Spedicato_2017, Zou2019}.
Further, it stands to reason that taking advantage of the graph's
theoretical representation of the transition process to model the
relationships between the event processes is both natural and
beneficial. Nevertheless, an explicit formalization of the transition
networks has not been leveraged nor has it been applied to the
learning process. TNA offers a robust framework and a
theoretical grounding for such modeling, expanding over existing
approaches by introducing several functions, e.g., validation methods. Moreover, it offers an R package \cite{Anonymous_2024} as well as a set of step-by-step tutorials to aid researchers in implementing TNA \cite{Saqr2025-vs,Lopez-Pernas2025-bi,Saqr2025-kn}.

While TNA may be viewed as a combination of two existing approaches ---and it may be partially so---, TNA offers more than the sum of its parts
(networks and SPM). Graph representation extends SPM by allowing
researchers to capture the role of an event in shaping the learning
process through centrality measures. For instance, the nodes that bridge
transitions (e.g., betweenness centrality), or the event that receives
most transitions (e.g., indegree centrality). Edge-level centralities
can also offer valuable insights about the transitions that undergirds
the learning process (e.g., edge centrality). Furthermore, through
community finding researchers can examine which events occur together
(e.g., patterns) that shape the structure of behaviors and
approaches. SPM extends the capabilities of network analysis and offers
clustering into typical patterns of transitions (network clusters).
These clusters can be used ---and, in fact, have been commonly used--- to reflect patterns of
students\textquotesingle{} behavior (tactics and strategies)
\cite{Matcha_Gašević_Ahmad_Uzir_Jovanović_Pardo_Maldonado-Mahauad_Pérez-Sanagustín_2019,Saint_Gaševic_Matcha_Uzir_Pardo_2020}. Further inclusion of covariates into the TNA model helps explain why certain patterns emerged or which clusters are likely to be associated with a variable beyond the interaction data (e.g., performance or context).
More importantly, TNA adds functionalities to both methods. The inclusion of bootstrapping in particular helps establish model stability and eliminate spurious edges using a rigorous method. Furthermore, pattern mining in TNA (dyads, triads, and cliques) offers a useful method for finding interdependent and connected behavioral elements. 

Lastly, while our case study was based on collaborative learning, TNA can be used to model a vast array of states, events, and processes that involve learning and
learners or ---dare we say--- any temporal process at large. For instance, TNA 
can be used to model transitions of attitude, roles, regulations states,
motivations, conditions, or any discrete event that unfolds in time where a transition analysis is deemed informative.

Currently, the existing development in TNA is expanding the validation
routines and, in particular, expanding bootstrapping, building null
models that are appropriate for directed weighted networks. While not
demonstrated due to space limitations, TNA supports the disparity
algorithm as a null model to validate the network structure. We have
also developed methods for estimating TNA graphs 
based on frequency which are suitable where Markov assumptions do not hold. A longitudinal TNA model
is also being developed with a range of visualization and statistical
metrics for the longitudinal process. Besides, we are currently improving and expanding
the existing R package to include more functions for estimating and
visualizing different networks with different options. Given the recency
of the methods, several tests and experiments are being performed to
test the functionalities and applications using simulated data, e.g., the
value and meaning of different centrality measures in a transition
process, such as diffusion centrality.

Of course, not all network analysis methods are transferable to TNA. Centrality measures ---and, in fact, all network measures--- are context-sensitive. In TNA, out-strength centrality is always one (given that probabilities sum to 1) for all nodes and is, therefore, not useful in non-pruned networks. Similarly, non-directed centralities and centralities that are built for certain contexts, e.g., information centrality are not applicable in TNA. We also need to caution researchers before transferring concepts or ideas from SNA or other graph measures that they should use an appropriate theory-based measure with the right calculation methods.

As is the case with all methods, TNA is not without limitations. Reliable estimation of transition
probabilities requires appropriate sample sizes, especially when some
transitions are rare. As is the case with Markov models, we assume
that the transition probabilities do not change over time and that the
transitions only have a first-order dependency on previous states. As
the number of states in a model increases, the complexity of the
transition matrix may increase exponentially. Also, with complex
transition matrices interpreting the model and its transition dynamics
may become difficult. In the case of mixed Markov models, the estimation
task can be computationally intensive. Furthermore, TNA may produce
fully connected networks that may not be useful in cases where the
transition processes lack enough variability; however, this scenario is
rather rare. In some situations, Markov models fail to converge and have
estimation problems that make it difficult to rely on the methods.
Additionally, Markov models may be difficult with continuous-time
processes and are sensitive to initial conditions.

%\bibliographystyle{unsrt}  
%\bibliography{sample-base}

\end{document}